# Design of the Readout Electronics for the Qualification Model of DAMPE BGO Calorimeter

Changqing Feng, Deliang Zhang, Junbin Zhang, Shanshan Gao, Di Yang, Yunlong Zhang,

Shubin Liu, Qi An

*Abstract*–The DAMPE (DArk Matter Particle Explorer) is a scientific satellite being developed in China, aimed at cosmic ray study, gamma ray astronomy, and searching for the clue of dark matter particles, with a planned mission period of more than 3 years and an orbit altitude of about 500 km. The BGO Calorimeter, which consists of 308 BGO (Bismuth Germanate Oxid) crystal bars, 616 PMTs (photomultiplier tubes) and 1848 dynode signals, has approximately 32 radiation lengths. It is a crucial sub-detector of the DAMPE payload, with the functions of precisely measuring the energy of cosmic particles from 5 GeV to 10TeV, distinguishing positrons/electrons and gamma rays from hadron background, and providing trigger information for the whole DAMPE payload. The dynamic range for a single BGO crystal is about $2\times10^5$ and there are 1848 detector signals in total. To build such an instrument in space, the major design challenges for the readout electronics come from the large dynamic range, the high integrity inside the very compact structure, the strict power supply budget and the long term reliability to survive the hush environment during launch and in orbit. Currently the DAMPE mission is in the end of QM (Qualification Model) stage. This paper presents a detailed description of the readout electronics for the BGO calorimeter.

*Index Terms*—DAMPE, Calorimeter, Dark Matter, Cosmic Ray, Areospace Electronics

## I. INTRODUCTION

Observing the high energy particles in space is regarded to be a hopeful way for the indirect search for dark matter. Several experiments have already been carried out, such as the ATIC Antarctic Baloon [1], the Fermi/LAT satellite [2], the PAMELA satellite [3] and the AMS02 [4] experiment. In China, a scientific satellite named DAMPE (DArk Matter Particle Explorer) has been proposed [5], with a 500 km orbit altitude and a planned mission period of more than 3 years. The major scientific objectives of DAMPE mission are cosmic ray study, gamma ray astronomy, and searching for the clue of dark matter by measuring the spectra of high energy electron/positron and gamma rays.

Manuscript received June 16, 2014. This work was supported by the Strategic Priority Research Program on Space Science of Chinese Academy of Sciences (Grant No. XDA04040202-4), and the National Basic Research Program (973 Program) of China (Grant No. 2010CB833002).

The authors are with Department of Modern Physics, University of Science and Technology of China, and State Key Laboratory of Particle Detection and Electronics, USTC, Hefei, 230026, P.R. China ( telephone: +86-551-63600408, e-mail: fengcq@ustc.edu.cn; dlzhang@mail.ustc.edu.cn; junbin@mail.ustc.edu.cn; shawn@mail.ustc.edu.cn; dyg87@mail.ustc.edu.cn; ylzhang@ustc.edu.cn; liushb@ustc.edu.cn; anqi@ustc.edu.cn ).

As illustrated in Figure 1, the DAMPE payload mainly consists of a Plastic Scintillation Detector (PSD), a Silicon Tracker (STK), a BGO Calorimeter (BGO) and a Neutron Detector (ND). The BGO Calorimeter, which is composed of 308 BGO (Bismuth Germanate Oxid) crystal bars with the size of 2.5 cm×2.5 cm×60 cm for each bar, is a crucial sub-detector for measuring the energy of cosmic particles, distinguishing positrons/electrons and gamma rays from hadron background, and providing trigger information. All the BGO bars are stacked in 14 layers, with 22 BGO crystals in each layer, to achieve nearly 32 radiation lengths. The BGO bars of every two adjacent layers are oriented perpendicularly to provide an X-Y measure of the particle hit position.

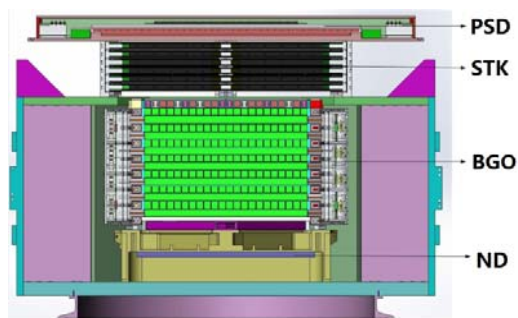

Figure 1. Architecture of DAMPE payload

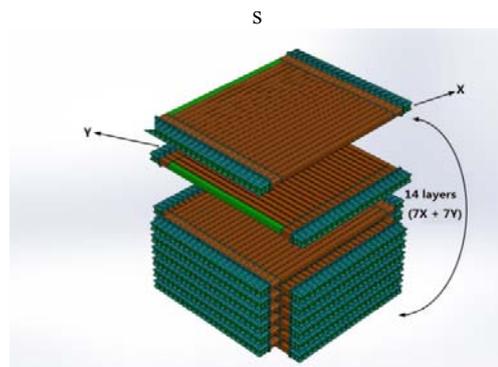

Figure 2. Arrangement for crystal bars of BGO Calorimeter

## II. REQUIREMENTS FROM THE DETECTOR

According to physical simulation, for a 10 TeV high energy electron, the maximum deposited energy in one bar is about 2 TeV. The deposited energy from a minimum ionizing particle (MIP) passing through the 2.5 cm thick BGO bar is about 23 MeV. To precisely sample the cosmic ray shower profile, which is regarded to be critical for hadron background rejection, a lower energy limit less than 0.5 MIPs is needed. Therefore a measurement range from 11 MeV (about 0.5 MIPs) to 2 TeV is required for each BGO bar, which is equal to about $2\times10^5$ and exceeds the capability of one single electronics channel, which can only be assured to about $10^3$, most of the time.

In order to achieve this large range, a three-dynode (2, 5 and 8) scheme for the PMT base design is proposed [6]. As shown in Figure 3, each crystal is viewed from both sides by two Hamamatsu R5610A-01 PMTs (photomultiplier tubes). An attenuating filter is inserted between each PMT photocathode window and its corresponding BGO bar, to adjust the PMT output response to the deposited energy. Each PMT send out 3 dynode signals to the readout electronics respectively. Thus there are 1848 readout channels for the 616 PMTs in total.

With proper PMT base design, the relative gain from dynode 2 to 8 is guaranteed to be greater than $10^3$ with good linearity, under ordinary high voltage. The basic concept is use uniform readout channels to couple different dynodes, and the relative gain among the three dynodes can be utilized to supplement the dynamic limitation of a single readout channel.

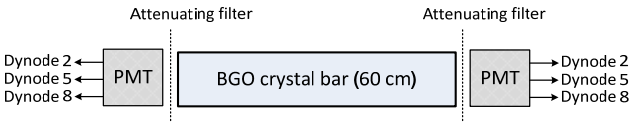

Figure 3. Readout scheme of BGO detector

As demonstrated in Figure 4, the dynode 8, with highest gain, stands for the low energy range, while dynode 2, with lowest gain, stands for the high energy range, and dynode 5 stands for the medial energy range. Here we can make some assumptions: the signal from dynode 8 is adjusted to 100 fC for a single MIPs energy deposition; the relative gains from dynode 2 to 5 and from dynode 5 to 8 are both is about 50 times; the lower sensitivity limit for the Pre-Amplifier is less than 30fC and the saturation level is about 12 pC (refer to the VA160 ASIC which will be introduced in the following chapter). Then we can easily deduce that a dynamic range from 0.5 MIPs to $1.1\times10^5$ MIPs can be achieved.

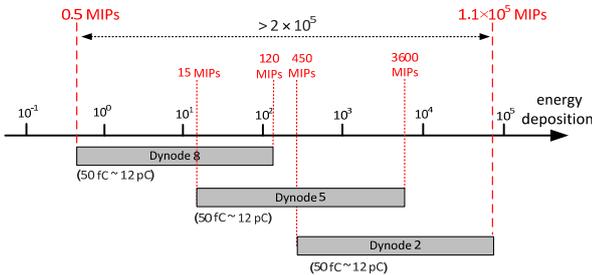

Figure 4. Design concept to achieve the large dynamic range

Besides performing charge measurement with high precision and large dynamic range, another important function of the BGO calorimeter is providing trigger information for the whole DAMPE payload. According to the physical design scheme, the trigger information is derived from top four layers (Layer 1, 2, 3, 4) and bottom four layers (Layer 11, 12, 13, 14) among the fourteen layers of BGO detector. The signal that carries trigger information is named "hit" signal (sometimes also called sub-trigger). If one or more BGO bars in a certain layer are hit by high energy particles, a digital pulse will be generated and sent to the PDPU (Payload Data Process Unit) crate. The PDPU receives all the hit signals from different layers, processes them and generates a global trigger signal for the DAMPE payload.

In the BGO detector, only dynode 5 and dynode 8 channels, which stand for low and medial energy range, are used to generate "hit" signals. The trigger threshold in the BGO calorimeter should be programmable in orbit, in order to realize flexible trigger logic.

III. SYSTEM ARCHITECTURE OF THE READOUT ELECTRONICS

The architecture of the readout electronics for BGO calorimeter is presented in Figure 5. The BGO detector has four vertical sides, each with 154 PMTs and 462 (154×3) signal channels. On each side, there is a group of FEEs to take charge of the readout task for the detector signals.

Besides, there are four HV crates and four DC-DC crates to provide high voltage supply for the PMT and to provide low voltage for the FEEs, respectively. One HV crate and one DC-DC crate are placed on each side, to provide high voltage or low voltage power supply for the nearby PMTs and FEEs.

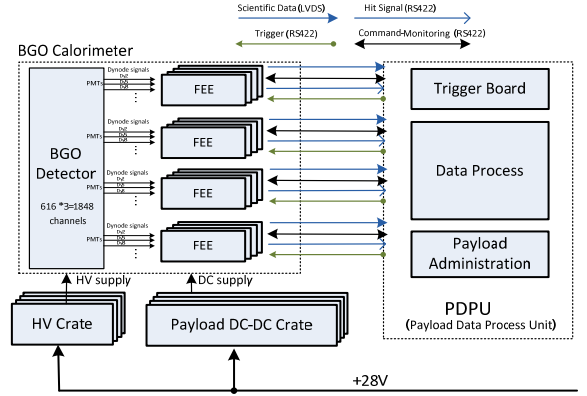

Figure 5. Architecture of the readout electronics for BGO calorimeter

The FEE receives the current pulse from PMT dynode, integrate the charge, and digitize it under the control of trigger signal. The measured result will be packed in the FEE, and transmitted to PDPU by a user-defined serial protocol. The protocol is based on LVDS signal level and the clock frequency is 20 MHz.

For the FEEs who are in charge of the top four and bottom four layers of the detector, signals from dynode 5 and dynode 8 will be discriminated to generate the original hit signal. The original hit signal is processed in the FEE to form a RS422 negative pulse with the width of 1000 ns, and sent to PDPU crate.

A Trigger Board in the PDPU crate receives all the hit signals from BGO calorimeter, and determines whether to generate a global trigger signal. The minimum time interval between two adjacent trigger pulses is about 3 ms, thus the signal processing and data transmission process for the readout electronics must be finished within 0.8 ms, to assure that the PDPU has adequate time to pack all the scientific data in a CCSDS 732.0-B-2 [7] format and write them into a nonvolatile mass storage.

The command from PDPU to FEEs and the status information from FEEs to PDPU are both transmitted by a RS422-based 115.2Kbps half-duplex serial bus (also called command-monitoring bus).

## IV. CHALLENGES AND SOLUTIONS

The large amount of detector components and signal channels, as well as large dynamic range requirements and the hush space environment, greatly challenges the design of the BGO calorimeter readout electronics.

### A. High Integration Level

The first challenge comes from the strict weight and size budget for the satellite, which imposes very stringent constraints on the mechanical design. Thus effective space for containing the electronics boards is limited as well, which results in confined area for the PCB and cable installation. This leads to a design strategy to reduce the complexity of FEE modules, and to increase the integration level as much as possible.

Figure 6 illustrates the system configuration for the BGO detector and its readout electronics. Three types of FEE modules are designed, named FEE-A, FEE -B and FEE -C. FEE-A and FEE-B, are both in charge of the 132 input channels in 2 layers; while FEE-C, with 66 input channels, only readout 1 layers. Besides precisely measuring the charge of PMT signals, the FEE-A module (for reading out Layer 1, 2, 4, 11, 12, 13 and 14) needs to provide hit signal to PDPU crate as well.

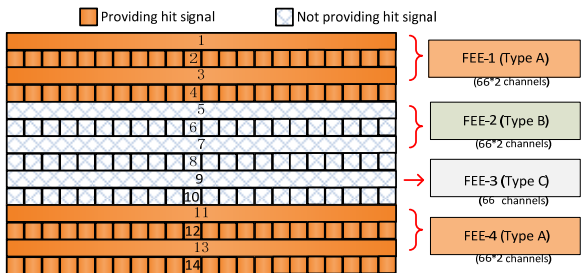

Figure 6. Configuration for the BGO detector and FEEs

Totally there are 16 FEE modules used for the BGO detector, with 4 modules assembled on each side. All the FEE PCB boards are designed within the size of roughly 11 cm × 38 cm, and installed vertically and very closely to the detector, to adapt to the size limitation of the supporting structure.

### B. Low Power Consumption

The second challenge comes from the strict power budget. It becomes a critical issue in the BGO calorimeter not only because of the limited capacity of the satellite's solar battery array, but also resulting from the simplified thermal design mainly due to weight and power reduction strategy. As the FEEs and detector components (PMTs and BGO bars) are integrated closely together, the operating temperature of the detector is directly influenced by the heat comes out from the FEE boards. To properly restrict the operating temperature of PMTs is considered to be a useful way to assure the lifetime of PMTs and to decrease the dark noise. This can be significantly achieved by cutting down the power dissipation of FEEs and increase their thermal conductivity to the outer shielding panel.

Furthermore, as the photon yield of the BGO crystal and the gain of PMT were both proved to have a negative temperature coefficient [8], the performance of the BGO detector can also benefit from reducing the power dissipation of FEE boards, because this makes the temperature field inside the detector more symmetrical and uniform.

Though there are nearly two thousand signal channels need to be read out, finally the power budget for all the FEEs is less than 40W. To overcome the challenges, two types of 32-channel ASIC, named VA160 and VATA160, with the power dissipation of about 150mW and 200mW respectively, are chosen as the front end chips to read the signal from PMT, and to provide fast trigger information.

### C. Reducing the Crosstalk

The third challenge is the potential crosstalk among the signal channels, due to high density cable assembly and PCB layout, and especially caused by the large dynamic range of the signals.

For BGO calorimeter, the relative gain from PMT dynode 2 to dynode 8 is greater than $10^3$, which makes the crosstalk a severe problem. Even we optimistically assume that the crosstalk coefficient between two adjacent readout channels is as little as 0.01%, which can be ignored by most ordinary applications, the maximum crosstalk in BGO calorimeter can reach to 10% from dynode 8 to another dynode 2. What is worse, dynode 2 stands for the particles in highest energy range, which are rare events and closely related to the major scientific objective of DAMPE mission.

In order to overcome this, the first measure is to use high quality shielded twisted cables to transmit PMT signals, besides careful PMT base layout.

The second measure is using different VA160/VATA160 ASIC chips to receive different dynode signals separately. In the FEE board, ASIC chips dedicated for dynode 2, 5 and 8 are orderly mounted, with enough distance from one to another. Therefore the dynode 2 chip is farthest away from dynode 8 chip, so that the crosstalk between them can finally be ignored. These measures, as well as proper PCB layout method, make the large dynamic range of $2*10^5$ finally be guaranteed, which was proved by LED test [6] and CERN beam test in Oct. 2012 [5].

### D. High Reliability in Space Environment

There are some other particular requirements derived from hash space environment, such as good radiation tolerance, proper thermal control, etc.

In order to estimate the performances of the key devices (e.g. the VA160 and VATA160 ASICs and the FPGA chips) in radiative space environment, both single event effect (SEE) and total ionizing dose (TID) tests are conducted, and some SEU and latch-up protection methods are adopted, based on the test results.

Meanwhile, thermal simulation for the FEE modules was performed to assure that all devices will operate in a proper temperature condition.

The methodology of redundant design is adapted in the BGO calorimeter to increase the system reliability. Firstly BGO crystal is viewed on both ends so that the deposited energy in one single bar can be measured by two PMTs and their corresponding FEEs, which forms a twofold redundant. In the FEE design it is difficult to use low level redundancy, mainly because of the limitation area for PCB layout. But all the power supply cables and all the interfaces with PDPU are designed with twofold hot redundancy.

The FEE also complies with other design policies for aerospace electronics, such as ESD protecting, EMC design, Grade 1 derating, etc., to assure the long term reliability.

## V. Design of the FEE module

The BGO calorimeter consists of 8 FEE-A modules, 4 FEE-B modules and 4 FEE-C modules, with 4 modules assembled on each side.

The three types of FEEs are designed with the same scheme, except that their numbers of input channel are different, and FEE-A need to send out hit signals. This paper mainly describes the FEE-A module, which is most complicated in the three types.

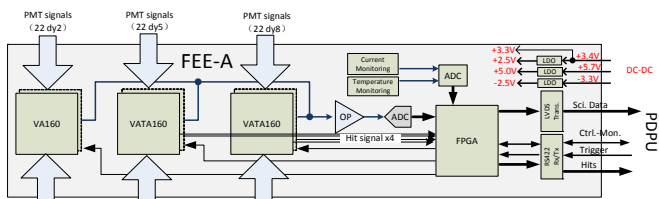

Figure 7．Block Diagram of FEE-A

The block diagram of FEE-A is given in Figure 7. Its major parts will be introduced below.

### A. Detector Signal Input and ESD Protecting

In order to avoid the serious crosstalk between strong signals (dynode 8) and weak signals (dynode 2), cables with shielding layer is a must. Usually a high quality coaxial cable is a best candidate, but it is more rigid and the connector occupies too much PCB area, which is not endurable for the FEE design because of the large amount of input channels and limited PCB size.

Finally a kind of space-qualified shielded twisted pair cable and Nicomatic CMM high density connector are applied. The twisted pair is used as signal and ground (backflow path in fact), and the shielding metal mesh is soldered to ground only on the end of PMT base. Near the FEE board, the two wires of the twisted pair are inserted into the CMM female connector by a kind of crimp pin, while the male connector is mounted on FEE board directly.

Each connector houses the 22 twisted pairs from the same kind of dynodes from one layer, and feed them to a single ASIC chip. Although there exists tiny crosstalk among the signals in the same connector, it will not become an important problem because all the signals belong the same energy range.

An ESD test for VA160 and VATA160 (bonded in a CQFP128 package) conducted in August 2013 indicated that their ESD level for digital IO pins is greater than 2000V, but the ESD level for analog input pins is just about 500V, in Human Body Model.

In case of Electro-Static Discharge (ESD) and possible overvoltage "spark" from high voltage components in PMT base, a clamping circuit is designed for protecting the front-end ASICs. The diagram is shown in Figure 8, which is composed of two diodes and a serial resistor. A Microsemi silicon diode array (1N5772) with SMT (Surface Mounted Technology) ceramic package is adapted. The ESD protecting level of 1N5772 is 8000V for direct contact in the IEC 61000-4-2 standard.

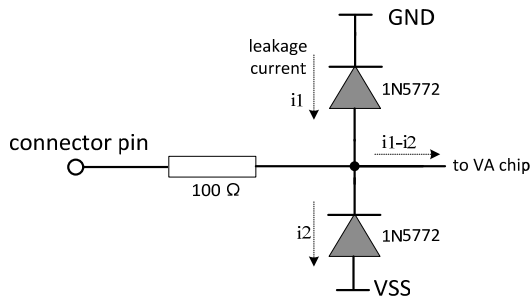

Figure 8．Block diagram of the protecting circuit

But this ESD protecting method may introduce a new problem, which comes from the leakage current from the reverse biased clamping diodes. The leakage are usually dozens of nA at room temperature, and will increase rapidly when the operating temperature rises, obeying the exponential function. Though for ordinary digital IOs this leakage current is not a problem, however for a charge sensitive amplifier with an internal feedback resistor with the value of several Mega ohms, it may cause serious error especially at high temperature.

However, it is lucky that usually a couple of diodes are used at the same time, then most part of the leakage will flow from one to another directly. As shown in Figure 8, only the residual (i1-i2) will flow to the VA chip.

In order to reduce the influence of this residual current, firstly a strict screening process was conducted for every 1N5772 components at room temperature. Only the diodes with low leakage currents (e.g. less than 60nA) and whose diode pairs have best leakage symmetry were submitted for board level assembly. Secondly all the QM FEE board were tested at 55℃, and a very few of 1N5772 components who affected the FEE performance were replaced. In the stage of

FM (Flight Model), we plan to screen the components more strictly, directly at the temperature of 55℃ or even higher.

*B. Charge Measurement*

The charge measurement function is performed by VA160 and VATA160 ASICs. There are 2 VA160 and 4 VATA160 used in a FEE-A module, 6 VA160 used in a FEE-B module and 3 VA160 used in a FEE-C module.

VA160 and VATA160 ASICs, optimized for the PMT dynode signals of DAMPE (Plastic Scintillation Detector and BGO detector), are both developed by IDEAS Inc. in Norway [9]. The VA160, which can be seen as an advanced version of the VA32HDR14.2 ASIC, has 32 analogue input channels for measuring the charge of PMT dynode signals, with a dynamic range from -3pC to 12pC and an INL error less than 2%.

The VATA160 can be seen as a combination of a VA part and TA part. The VA part is exactly the same as VA160, while the TA part is an advanced version of TA32CG ASIC and used to generate fast trigger ("hit" signals for BGO detector).

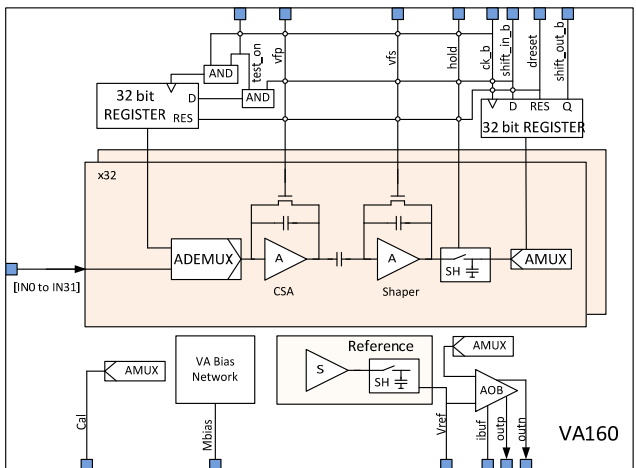

Figure 9． Block diagram of VA160 ASIC

The block diagram of VA160 is illustrated in Figure 9. Each input channel is composed of a charge sensitive pre-amplifier (CSA), a CR-RC shaping amplifier and a sample-hold circuit (SH). Signal from PMT dynode is integrated by CSA and then shaped into a semi-Gaussian pulse with the peak time of about 1.8 us, while the signal peak is proportional to the input charge.

A Hold signal is used to sample all the analog channels at the same time. When Hold goes high, the analog switches in all the sample-hold circuits will be turned off, and the 32 shaping amplifier outputs will be stored in their capacitors and waits for being sent out sequentially in the form of differential current, under the control of a chain of shifter registers and an analog multiplexer. Beside the 32 input channels, there is a dummy channel inside the ASICs. The dummy channel is used as a reference for the output driver, which greatly reduce the common noise and temperature drifting.

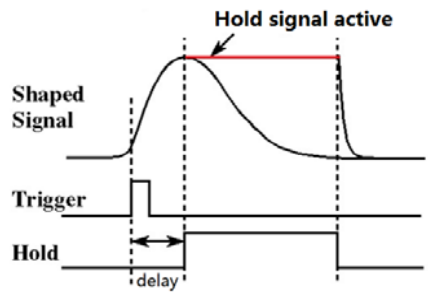

Figure 10． Time diagram of charge measurement process with VA160

The output current drivers of all VA160 and VATA160 chips in one FEE are wired together and led to a current-to-voltage circuit formed by operational amplifiers with the output range from about 0 to 5V, and digitized by an AD976A chip, which is a 16-bit ADC with an input range from -10 V to +10V but only need +5V single supply.

Theoretical analysis and test results both suggest that there is an obvious negative correlation between the noise and the output impedance of the detector. As the capacitance of signal cable is fixed (mainly depending on the cable parameters and its length), the only measure is to constrain the output capacitance and resistance of PMT base. Finally the value for all components in PMT base are carefully chosen and proved by test results, and a 10 k ohm output series resistor and a 1 nF output series capacitor are used.

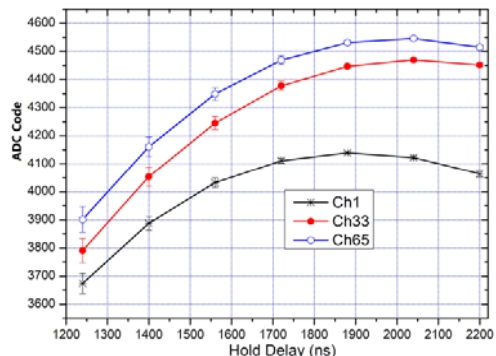

Figure 11． Measurement result (ADC code) versus Hold delay

The delay time of Hold signal is another critical issue for the measurement error. A test result is illustrated in Figure 11, in which the relation between measured charge (in ADC code) and Hold delay are shown. The curves just seem like top part of a CR-RC semi-Gaussian waveform and the roof is relatively smooth. We can see that the hold delay and jitter within +/-150 around peak may cause an error of about 2%.

In FEE module, the Hold signal is derived from the system Trigger, and the delay time can be configured in orbit, in order to catch the peak of shaper output in different trigger modes.

*C. Hit Signal Generation*

The FEE-A module needs to send out hit signals and this function is performed by VATA160 ASIC.

The block diagram of VATA160 is shown in Figure 12.

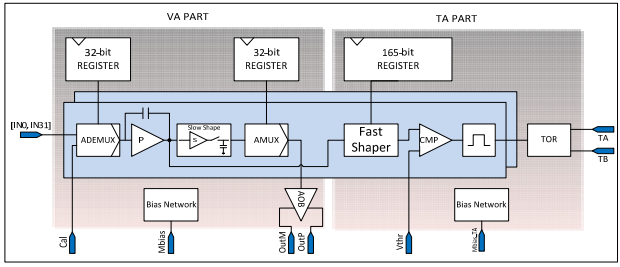

Figure 12. Block diagram of VATA160 ASIC

The VATA160 is composed of VA part and TA part. There are 32 channels in the TA part as well. Each channel contains a faster shaper and a comparator. The integrated pulse from the CSA (of VA part) will firstly be shaped into a fast narrow pulse and then discriminated by a comparator, to generate a digital signal. All the 32 outputs of TA part are internally OR'ed together, thus the output signal (in the form of differential current, named TA and TB) is shared by all input channels, which greatly simplifies the system design.

Each FEE-A is in charge of two BGO layers, while each layer needs two hit channels: the dynode 5 hit and dynode 8 hit. Therefore 4 VATA160 chips are applied in one FEE.

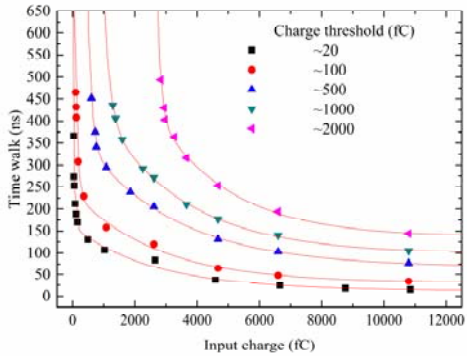

Figure 13. Time walk versus input charge, for different thresholds

On one hand, the signal from BGO detector is not fast, because of the 300 ns fluorescence decay time of BGO crystal. On the other hand, due to the parameter dispersion of BGO crystals and PMTs, it's hard to make the detector units from one layer have a uniform response to the energy deposition. Thus an important issue for VATA160 is the time walk of hit signals, which may cause error for charge measurement, as indicated in Figure 11.

A detailed test was conducted [], using a LED source to illuminate the PMT. The LED is driven by a BGO-like current pulse (an exponential curve with a 300ns decay time). The test result is shown in Figure 13, which gives the relationship between time walk and input charge at different thresholds. From the picture we can draw a conclusion that a lower threshold-to-signal ratio leads to smaller time walk error. For a threshold-to-signal ratio less than 1/2, the time walk is about 300 ns (can also be seen as +/-150ns), which is endurable.

An import solution is to control the uniformity of the detector in one layer during component test and assembly procedures. It is beneficial both from reducing time walk and improving the trigger efficiency. Though it was a complex and difficult job, finally a less than 20% uniformity deviation (in sigma) was achieved for each layer in the QM stage.

### D. Calibration Circuits

In order to monitor the performances of electronics in orbit, a calibration circuit is designed, utilizing the calibration function of VA160 and VATA160 ASICs. This calibration circuit can also be used to test the hit generation function of VATA160.

As shown in Figure 14, the calibration circuit is mainly composed of an analog switch and a 5 k ohm resistor. When the analog switch is turned on, the circuit will generate a step pulse (from 0 to $V_{cal}$), and a current pulse is injected into the ASIC chip through an external 10 pF capacitor ($C_{cal}$).

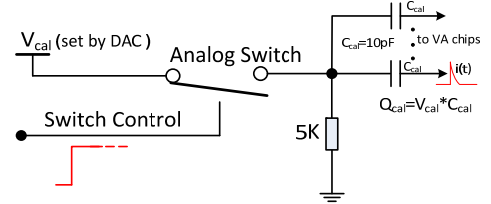

Figure 14. Block diagram of the calibration circuit

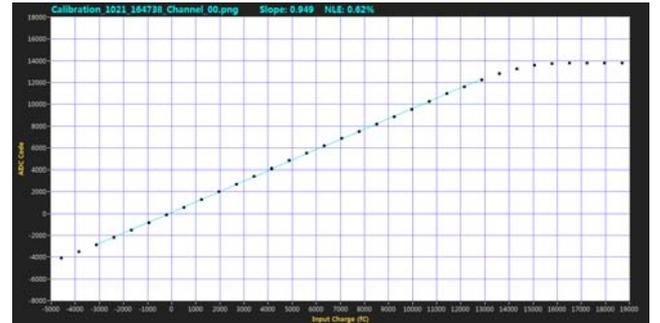

Figure 15. Calibration scan curve of a VA160 input channel

A DAC and an op-amplifier are used to provide the calibration voltage ($V_{cal}$), and the calibration charge injected to each VA chip is the product of $V_{cal}$ and $C_{cal}$. In the VA ASIC, an analog de-multiplexer is used to determine which input channel is enabled for calibration. By sequentially select the input channels and gradually increasing the calibration voltage, the scan curve of all channels in a FEE can be obtained. One calibration curve is illustrated in Figure 15, and in the figure the nonlinearity from 0 to 13 pC is less than 1%, which is better than the design specification.

### E. House-Keeping Circuits

Each FEE module has 2 current monitoring channels, for the positive (+2.5V) and negative (-2.5V) supply current of all VA160/VATA160 chips, respectively. The current monitoring circuits are composed of resistors (1/4 ohm), op-amplifiers, and a multi-channel ADC shared by the thermal monitoring circuit.

Four thermal resistors are used for each FEE, among them three are used to obtain the temperature on BGO crystals and the other one is used to obtain the temperature on one of the VA160/VATA160 packages.

Besides the current temperature monitoring function, an array of status register (32 bytes) in the FPGA logic of FEE is sent to PDPU as well. PDPU will inquiry temperature and FPGA status every 16 seconds, while inquiry current every second.

*F. Radiation Protecting*

Three kinds of radiation effects are taken into consideration for the FEE design, including the TID (Total Dose) effect, the SEU (Single Event Upset) effect and the SEL effect.

The DAMPE satellite will operate in a near earth orbit with the height of 500 km. According to simulation, with 3 mm Al shielding, the total dose in three year is less than 3 krad. To ensure an adequate RDM (Radiation Design Margin), the design specification for BGO electronics is 20 krad, which can be achieved by most of the ordinary components.

For the BGO FEE modules, the detector with a weigh of about 800 kg (mainly contributed by BGO crystals) is an ideal shielding for about $2\pi$ solid angle. The other $2\pi$ solid angle is mainly shielded by the BGO envelope panels, other crates and supporting structures, with a total thickness at least 3.6 mm, which provide a good condition for TID protecting.

Several total dose tests were carried out in the end of 2013 to study the TID tolerance of VA160 and VATA160 ASICs, using $^{60}$Co gamma sources with an activity order of about $10^4$ Curie. Test results showed that the electric performances and the supply currents of both ASICs remain rather stable, with the total dose greater than 30 k rad and a dose rate of about 5 rad/s.

The SEU protecting is also taken into account for sequential logic and memories. Firstly TMR (Triple Modular Redundancy) method is used for the 165 bit configuration registers of VATA160, by IDEAS. For the FEE design, APA600 and APA300 FPGA, from Actel flash-based ProASIC Plus (APA) family, are chosen to implement the control logic of FEE.

The APA FPGA uses the floating gate to store the programming information of the logic elements, which can be regarded immune to SEU, which also proved by several heavy ion beam tests from 2012 to 2013. Test results also showed that the APA300 and APA600 is SEL immune.

However, the registers implemented from logic tiles and RAM blocks are still SEU sensitive, just like other SRAM-based FPGAs. In order to overcome this problem, several design methodologies were adapted and verified by heavy ion beam tests, such as TMR and CRC (Cyclic Redundancy Check), etc., whose effectiveness have already been verified by heavy ion beam tests [11].

Heavy ion beam tests showed that VA160 and VATA160 is sensitive to SEL, with the threshold of about 21 MeV*cm$^2$/mg and 11 MeV*cm$^2$/mg. By using the CREAM96 program, the SEL rates are calculated to be: 2.7E-5 device$^{-1}$*day$^{-1}$ for VA160 and 8.1E-5 device$^{-1}$*day$^{-1}$ for VATA160, which cannot be negligible.

A protecting solution is proposed and its principle is shown in Figure 16. The supply current of VA160/VATA160 ASICs are sampled by the house-keeping circuit, in a period of 100 us. If the currents exceed the preset threshold, the FPGA will send control signals to the two regulators (LDO), to disable the power supply of VA160/TA160. After about 30 ms, when the SEL disappears, the FPGA will enable the power supply again. A laser-pulse experiment is used to verify this solution. In the experiments, a laser-pulse is used to induce SEL events in the ASICs. Test results showed that every SEL events can be detected and recovered.

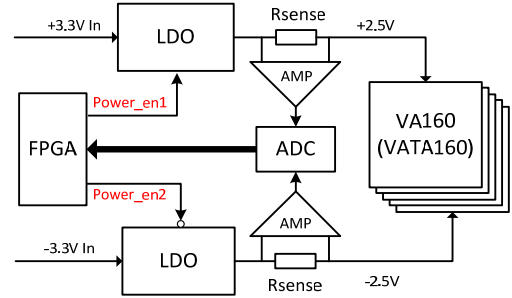

Figure 16. Block diagram of SEL protecting solution

VI. MASS PRODUCTION AND ENVIRONMENTAL TESTS

At the end of 2013, twenty BGO FEE QM modules, including four spare modules, were produced in a factory, complying with aerospace quality control regulations. All the modules were tested with a strict procedure, and passed the screening procedure with 16.5 cycles from -45℃ to 75℃ under ordinary pressure.

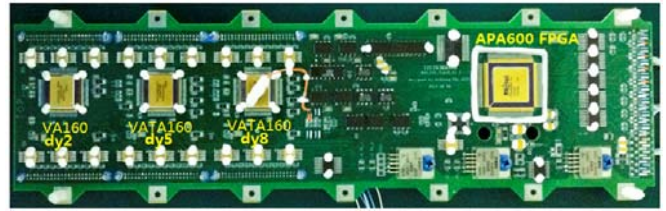

Figure 17. Photograph of the QM FEE board (Type A)

Figure 17 is the photograph of a FEE-A module. There are 2 VA160 and 4 VATA160 ASIC chips mounted on each board. Three ASICs are mounted on top side while the other three ones are mounted on bottom side. PMT signals from dynode 2, 5 and 8 are sent to the three ASIC (top or bottom side) from left to right respectively. FEE-B possesses six VA160, with three chips on each side, while FEE-C only has three VA160 on top side.

At last 16 FEE modules were installed with the detector at the beginning of 2014. A serial of environmental tests were successfully performed for the BGO calorimeter, including the EMC (Electromagnetic Compatibility) test, the vibration test, the thermal cycling test, and the thermal-vacuum tests. All the test result showed that the electronics satisfied the system requirements.

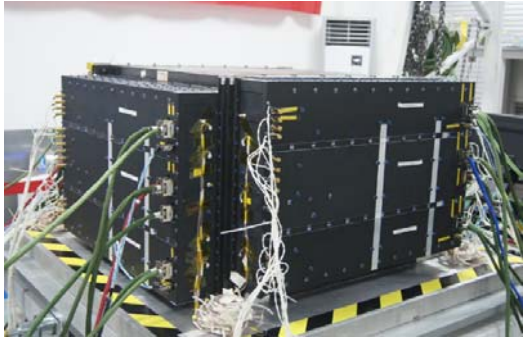

Figure 18．Photograph of the BGO calorimeter qualification model after assembly

Currently the BGO calorimeter is under integration tests with other sub-detectors, and a beam test will be carried out at CERN, in the autumn of 2014.

ACKNOWLEDGMENT

We would like to give thanks to Dr. Jianhua Guo, Mr Yiming Hu, Mr Dengyi Chen, Dr. Fei Zhang and Dr. Jie Kong. They discussed with us and gave many useful suggestions. We also would like to thank all the people from DAMPE collaboration who helped make this work possible.